%% file: main.tex
\documentclass[sigconf]{acmart}

\acmConference[WSDM 2022]{WSDM 2022: The 15th ACM International Conference on Web Search and Data Mining}{July 11--15, 2022}{Virtual, Online}

\setcopyright{none}
\renewcommand\footnotetextcopyrightpermission[1]{}
\usepackage{lineno,hyperref}

\usepackage{multirow}
\usepackage{makecell}
\usepackage[utf8]{inputenc}

\usepackage{listings}
\usepackage{xcolor}
\usepackage{subcaption}
\usepackage{mathtools}
\usepackage{amsmath}
\usepackage{footnote}
\usepackage{xspace}
\usepackage{verbatim}
\usepackage{booktabs}
\usepackage{array}
\usepackage{enumitem}
\usepackage{tablefootnote}

\newcolumntype{H}{>{\setbox0=\hbox\bgroup}c<{\egroup}@{}}

\newcommand{\virgolette}[1]{``#1''}

\newcommand{\algoname}[1]{\textsc{{#1}}\xspace}

\newcommand{\toppop}{\algoname{TopPop}}
\newcommand{\itemknn}{\algoname{ItemKNN}}
\newcommand{\userknn}{\algoname{UserKNN}}
\newcommand{\puresvd}{\algoname{PureSVD}}

\newcommand{\rpbeta}{\algoname{RP$^3_\beta$}}
\newcommand{\palpha}{\algoname{P$^3_\alpha$}}
\newcommand{\slim}{\algoname{SLIM}}
\newcommand{\easer}{\algoname{EASE-R}}
\newcommand{\als}{\algoname{ALS}}
\newcommand{\multvae}{\algoname{MultVAE}}
\newcommand{\recvae}{\algoname{RecVAE}}

\newcommand{\warp}{\algoname{WARP}}
\newcommand{\koswarp}{\algoname{k-OS WARP}}

\makeatletter
\newcommand\footnoteref[1]{\protected@xdef\@thefnmark{\ref{#1}}\@footnotemark}
\makeatother

\begin{document}

\title{Multi-stage Ensemble Model for Cross-market Recommendation}

\author{Cesare Bernardis}
\email{cesare.bernardis@polimi.it}
\orcid{0000-0002-8972-0850}
\affiliation{%
  \institution{Politecnico di Milano}
  \city{Milano}
  \country{Italy}
}

\renewcommand{\shortauthors}{Cesare Bernardis}

\begin{abstract}
This paper describes the solution of our team PolimiRank for the WSDM Cup 2022 on cross-market recommendation.
The goal of the competition is to effectively exploit the information extracted from different markets to improve the ranking accuracy of recommendations on two target markets.
Our model consists in a multi-stage approach based on the combination of data belonging to different markets.
In the first stage, state-of-the-art recommenders are used to predict scores for user-item couples, which are ensembled in the following 2 stages, employing a simple linear combination and more powerful Gradient Boosting Decision Tree techniques.
Our team ranked 4th in the final leaderboard.
\end{abstract}


\maketitle

\input{00-introduction}
\input{01-dataset}
\input{02-model}

\input{03-experiments}
\input{10-conclusions}

\begin{acks}
To my supervisor, prof. Paolo Cremonesi, and all the RecSys Research Group at Politecnico di Milano for the support.
\end{acks}

\bibliographystyle{ACM-Reference-Format}
\bibliography{bibliography}

\end{document}

%% file: 00-introduction.tex
\section{Introduction}
\label{sec:introduction}

Online shopping has become part of everyday life in several countries.
Recommender systems play a crucial role in this environment, as they guide users in the exploration of huge catalogs of products.
These catalogs often overlap between different markets, and e-commerce companies have to deal with the recommendation of similar sets of items in different scenarios.
This allows sharing both experience and information across markets, with the risk to spread market-specific biases and impose trends of data-rich markets to others \cite{cross-market-1, cross-market-2}.
How to effectively exploit information from different markets to improve recommendation quality remains an open challenge, which is the focus of the WSDM Cup 2022 competition on cross-market recommendation.
In this report we describe the solution of our team PolimiRank, which ranked 4th in the final leaderboard.

\section{Problem formulation}
\label{sec:problem-formulation}

The goal of the WSDM Cup 2022 competition on cross-market recommendation is to re-rank a set of items for each given user, in order to achieve the highest ranking accuracy, expressed in terms of NDCG@10, on two target markets (t1, t2).
Competition data includes user ratings in the form of (userID, itemID, rating) for the two \emph{target} markets plus three additional \emph{source} markets (s1, s2, s3).
Sets of users among different markets are mutually disjoint by assumption, while the item sets overlap.

The evaluation is performed on a leave-one-out split with sampled sets of items to rank.
Practically, the organizers provide two sets of 100 items for each user in the target market, one for validation and one for test.
Among each of the 100 item sets, only 1 item was actually rated by the user (i.e., the positive item): the objective is to rank the positive item in the highest position.
The true positive item in the validation is included in the competition data, while the positive item in the test data is kept hidden and used for the estimation of the final leaderboard scores.
All the other available ratings are included in the training data.
A validation split in the same format is also provided for the three source markets.

Additionally, the competition data include a preprocessed version of the training data of each market.
Ratings were previously normalized in the 0-1 range and all the users and items with less then 5 ratings (and their respective ratings) were eliminated from the data.
Due to some undisclosed preprocessing steps performed during data generation, the ratings in the preprocessed version might differ from those in the \virgolette{complete} one, and do not have a rating value associated (all ratings have value 1).

%% file: 01-dataset.tex
\section{Datasets preparation}
\label{sec:dataset}

\begin{figure}[t]
    \centering
    \includegraphics[width=0.8\linewidth, keepaspectratio]{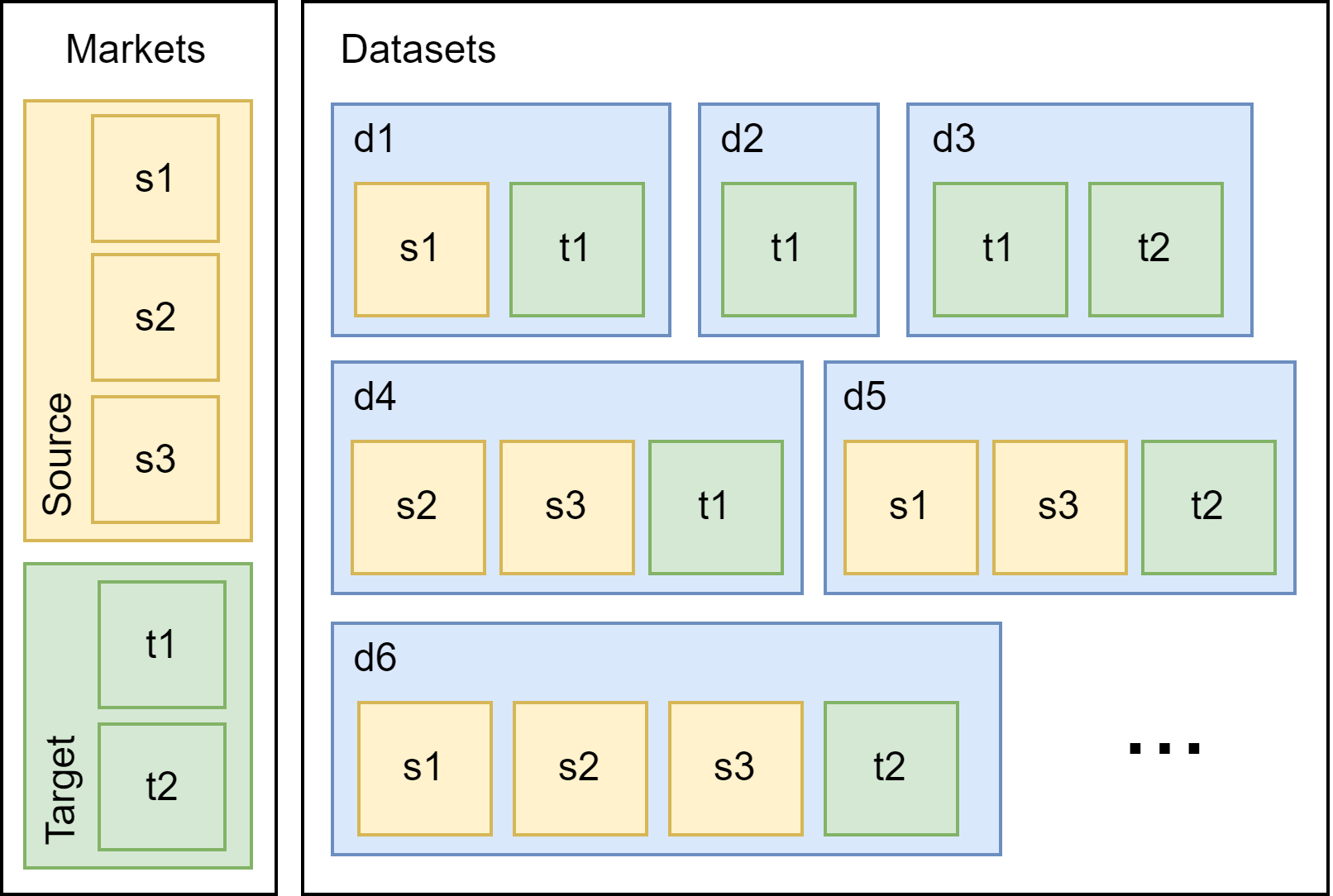}
    \caption{Target and, optionally, source markets data are combined to obtain datasets. Datasets are then used for training and evaluation of the models that compose our solution.}
    \label{fig:datasets}
\end{figure}

The underlying objective of the challenge is to push the competing teams to effectively exploit additional information coming from source markets, in order to enhance the accuracy of the rankings on the two target markets.
Our solution addresses this task by merging the information at data level, i.e., the interactions included in different markets are merged together in unique \emph{datasets}.
As mentioned in Section \ref{sec:problem-formulation}, the sets of users are assumed to be disjoint between markets, but the item sets overlap at least partially.
As a result, merging two markets consists in joining the user interactions of different markets in a single set of interactions.
Users coming from different markets are considered as new users participating in the same larger market that comprehends all the items of the markets.
In the following, we refer to \emph{dataset} as one of these fusions between data coming from one or more different markets.

We compute every combination of source and target markets, including target markets without source markets, to generate the datasets\footnote{Note that t1 and t2 are also used as datasets by themselves.}.
Since the goal is to rank items in the two target markets, we use only the combinations that included at least one target market, for a total of 24 final datasets.
An example of how markets are combined to obtain datasets is depicted in Figure \ref{fig:datasets}.
Both ratings and preprocessed training data are employed.
The unitary value of preprocessed data is substituted with 4 (i.e., the average positive rating) in order to unify the ratings scales.
Ratings for the same user-item couple are deduplicated by taking the average rating.

%% file: 02-model.tex
\section{Models}
\label{sec:models}

Our solution consists in a multi-stage model.
In the first stage, a set of state-of-the-art recommendation algorithms is employed to predict the scores for each dataset of the target market validation and test data involved.
These scores, plus some statistics of the dataset, are ensembled together, independently for each dataset, in a second stage.
As ensemble approaches we adopt three common Gradient Boosting Decision Tree (GBDT) implementations trained for the learning to rank task, and a simple linear combination of the recommenders predicted scores.
In the last stage, we use the most promising GBDT implementation to ensemble all the statistics and the scores predicted for each dataset, including the second stage ensemble predictions, in a unique final prediction.

\begin{figure}[t]
    \centering
    \includegraphics[width=0.8\linewidth, keepaspectratio]{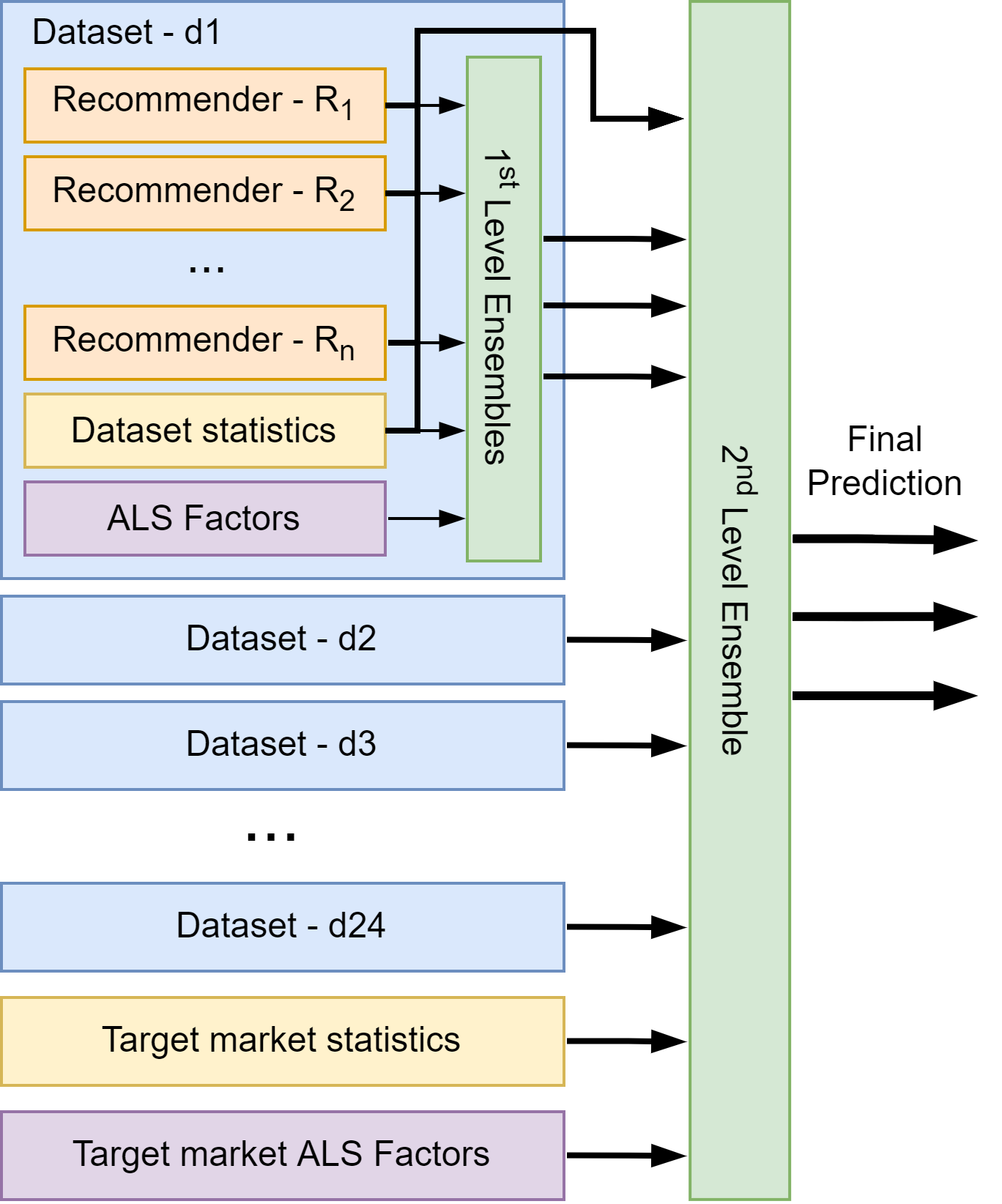}
    \caption{Graphical representation of the multi-stage model prediction on one target market.}
    \label{fig:model}
\end{figure}

In Figure \ref{fig:model} we propose a graphical representation of the structure of our multi-stage model.
The python code to reproduce our solution is publicly available on Github\footnote{https://github.com/cesarebernardis/WSDM-Cup-2022}.
In the following sections, we outline the details of each stage, describing in detail the employed models, and how training and validation are performed.

\subsection{Recommenders scores}
\label{sec:models:first-stage}

The first stage of our solution is executed independently for each dataset.
Some among the most common collaborative recommendation algorithms from literature are used to predict the scores for the validation and test data of the target markets that are included in each dataset.
The recommenders adopted are:

\begin{description}[leftmargin=15pt,itemindent=-5pt]
    \item [\toppop:] the predicted score of an item is proportional to its popularity (i.e., the number of users that rated the item).
    \item [\itemknn:] an item-based model that uses cosine to assess similarity between ratings vectors of items \cite{neighborhood-based}.
    \item [\userknn:] a user-based model that uses cosine to assess similarity between ratings vectors of users \cite{neighborhood-based}.
    \item [\palpha:]  a graph-based approach where scores are proportional to 3-step random-walks probabilities between users and items \cite{p3alpha}.
    \item [\rpbeta:] the model is equivalent to \palpha, but the final probability of each item is penalized by its popularity \cite{rp3beta}.
    \item [\puresvd:] a matrix factorization model based on Singular Value Decomposition \cite{puresvd}.
    \item [\warp:] a matrix factorization technique based on Weighted Approximate-Rank Pairwise loss optimization\footnote{\label{note1}We use the implementation in: https://making.lyst.com/lightfm/docs/index.html} \cite{warp}.
    \item [\koswarp:] a slight modification of \warp based on k-order statistic loss\footnoteref{note1} \cite{kos-warp}.
    \item [\slim:] an item-based model where the similarity matrix is obtained by solving a linear regression problem \cite{slim}.
    \item [\easer:] an item-based model recently proposed by Steck \cite{easer}.
    \item [\multvae:] a collaborative approach for implicit feedback based on variational autoencoders \cite{multvae}.
    \item [\recvae:] a combination of improvements over the \multvae autoencoder model \cite{recvae}.
    \item [\als:] a factorization model that uses an alternating-least-squares optimization process\footnote{We use the implementation in: https://implicit.readthedocs.io/} \cite{ials}.
\end{description}

Every recommender is trained on the training data presented in Section \ref{sec:problem-formulation}.
The hyperparameters of each algorithm are optimized independently for each dataset using the Bayesian Optimization approach provided in the \textit{scikit-optimize} library\footnote{https://scikit-optimize.github.io/stable/}, testing 50 configurations.

\subsection{Dataset-level ensemble}
\label{sec:models:second-stage}

In the second stage of our model, the scores predicted in the first stage are ensembled together dataset-wise.
The ensemble is performed using a simple linear combination of the scores, and with more powerful GBDT models.

\paragraph{Scores combination}
The first ensemble is performed computing a linear combination of the scores predicted in the first stage of the model, elevated to an exponent.
Before being combined, the scores are min-max normalized user-wise.
The ensemble is carried out independently for each dataset, using the validation and test data of the target market included in the dataset.
If both target markets are included, the ensemble is repeated separately for each target market.

Formally, given a recommendation algorithm $R$, and a user-item couple $(u, i)$, we define the normalized score of $R$ for that user-item couple as $\tilde{s}_R(u, i)$.
The final ensemble prediction $s(u, i)$ is computed as:
\begin{equation}
    \label{eq:ratings-combination}
    s(u, i) = \sum_{R} c_{R}(u) \cdot \tilde{s}_R(u, i) ^ {e_R(u)}
\end{equation}
where $c_R(u)$ and $e_R(u)$ are the recommender specific coefficient and exponents of each user.
To avoid excessive overfitting, we do not employ different coefficients and exponents for each user, but we group them by profile length, i.e., users that have a similar number of ratings in the dataset share the same $c_R(u)$ and $e_R(u)$ values.
In particular we divide users in 4 groups: 
\emph{short} profiles with length $< 5$;
\emph{quite short} profiles with $5 \leq$ length $< 8$;
\emph{quite long} profiles with $8 \leq$ length $< 12$;
\emph{long} profiles with length $\geq 12$;

To find the optimal values of $c_R(u)$ and $e_R(u)$ for all the recommenders, we treat them as hyperparameters, and we perform a hyperparameter optimization using the Optuna framework\footnote{https://optuna.readthedocs.io/en/stable/index.html}.
The goal of the optimization is to maximize the NDCG@10 of the evaluation performed on the validation data, where the items for each user are ranked accordingly to the decreasing value of the predicted score $s(u, i)$.
The same coefficients and exponents are finally used to predict the scores on the test data.

\paragraph{GBDT models}
We adopted 3 popular and successful implementations of GBDT: LightGBM\footnote{https://lightgbm.readthedocs.io/en/latest/}, XGBoost\footnote{https://xgboost.readthedocs.io/en/stable/} and CatBoost\footnote{https://catboost.ai/en/docs/}.
Thanks to their flexibility and robustness, they can easily adapt to different types of features, obtaining challenge-winning results\cite{recsys-challenge-1, recsys-challenge-2}.

The models are trained for the learning to rank task using a 5-fold cross-validation (CV) approach on the validation data of the target market included in the dataset, and they are used to predict the scores on the respective test data.
If both target markets are present in the dataset, training and prediction are performed separately on each target market.
The learning to rank task requires that samples, which represent a single user-item couple and contain the respective information under the form of features, are grouped together.
The goal of the model is to achieve the highest ranking accuracy (on average) among the groups.

Besides the scores of the recommendation algorithms obtained in the first stage, we added some supplementary features to each data sample:
\begin{description}[leftmargin=15pt,itemindent=-5pt]
    \item[Dataset statistics:] basic information about users and items that can be mined through the investigation of the available data, like the avarage rating of each user and item, the popularity of each item, and the number of ratings assigned by the user.
    \item[Latent factors:] we include the latent representations of users and items obtained through the factorization with 12 dimensions performed by the \als recommender. The representations are L2-normalized, and each latent factor corresponds to a new feature. The goal is to give the model the chance to recognize the user and the item involved in the sample it is scoring, but simply providing the respective identifiers might not be effective, since users and items appear in a limited number of samples and groups of validation and test data.
\end{description}

To train the model, the validation data is randomly divided in 5-folds, ensuring that each group appears entirely in only one fold.
Each fold in turn is used to validate the model trained on the data contained in the other 4 folds.
As a result, we obtain 5 models trained on different, but overlapping portions of data, and each of them is used to predict the scores on the fold that was not used for training, and on the test data. 
Multiple predictions on the same sample but from different models are averaged together to obtain a unique predicted score, that is used as feature in the next stage of the model.

Note that the validation of the model comprehends the handling of the early stopping of the GBDT models, when available, and the tuning of the hyperparameters.
In particular, the hyperparameter configuration chosen is the one that provides the highest NDCG@10, on average, on the 5 CV folds.

Interestingly, repeating the CV training and validation multiple times with folds differently composed allows reaching a higher generalization, improving the accuracy on the test data.
Therefore, we replicate the CV three times by changing the random seed used to split the sample groups among folds.
The final prediction score is computed as the average of the scores obtained in each repetition.

The whole training process is also repeated using different types of normalization on the recommendation scores used as features.
In particular, we use non-normalized scores, user-wise min-max normalized scores, and both the previous together, obtaining three different final predictions that are used in the last stage.

\subsection{Last level ensemble}
\label{sec:models:third-stage}

In the last stage of our model we perform a stacking ensemble using the same techniques described for the second stage in Section \ref{sec:models:second-stage}.

For what concerns the scores ensemble, it is performed exactly as previously explained, but instead of having different coefficients and exponents for each recommender, we have different parameters per dataset.
Also in this case, users with similar profile lengths share the same parameter values.
This type of ensemble was only used to break ties in the GBDT ensemble prediction, due to its lower accuracy performance (see Section \ref{sec:results}).

For the GBDT ensembles, we do not repeat the process for all the three implementations, but we adopt only the most promising GBDT model (i.e., LightGBM) to ensemble all the features and predictions obtained in the different datasets, and produce a single, unique final prediction.
For each target market, all the features obtained from the datasets that include that market are collected and concatenated, including:
\begin{itemize}
    \item The statistics of each single dataset;
    \item The statistics of the target market;
    \item The factorized representations of users and items (with 16 dimensions) obtained as described in Section \ref{sec:models:second-stage} on the target market dataset;
    \item The scores predicted by each recommender for each dataset;
    \item The scores predicted by the ensembles.
\end{itemize}
Training and prediction are performed as expressed in Section \ref{sec:models:second-stage}.
Also in this case, we adopt a 5-fold CV repeated three times, changing the random seed in the generation of the folds.

%% file: 03-experiments.tex
\section{Results}
\label{sec:results}

\input{tables/tab-gbdt-validation-results}
\input{tables/tab-scores-validation-results}

In Table \ref{tab:gbdt-validation-results} we report the average NDCG@10 obtained on the validation data by the best hyperparameter configuration on the 5-fold CV.
The results are divided by target market and dataset. 
We also show the results obtained with different normalization types to explore the effectiveness of cross-market recommendation, comparing the accuracy of the same models trained on different combinations of data coming from different markets.

Clearly, using non-normalized scores as features leads to higher ranking accuracy, while including both non-normalized and normalized scores typically worsen the performance with respect to the non-normalized version, even though the margin is quite small.

Looking at the different implementations of GDBT, they reach similar best results.
However, LightGBM performs more consistently across different datasets, proving to be the most robust approach.
Interestingly, CatBoost achieves a quite high best score, considering that normalized scores are employed.

Looking at the markets included in the datasets, we highlight that when predicting for target market t1, the accuracy generally benefits from the presence of data coming from source markets s2 and s3 in particular.
At the same time, data from target market t2 degrades the performance in most scenarios, as can be seen by comparing the datasets composed of the same combination of source markets, but with different target markets.
Concerning the prediction on the t2 target market, s1 is the most valuable market, while the contribution of the other target market data (t1) is unclear, as it leads to better or worse accuracy, depending on the market combination in the dataset.

In Table \ref{tab:scores-validation-results} we show the results obtained by the scores ensemble on the validation set of each dataset.
We report the NDCG@10 obtained with different profile lengths (see Section \ref{sec:models:second-stage}), in order to study the impact of the number of ratings provided by a user on the accuracy of the recommendations.
It is evident that having more ratings is detrimental for the recommendation performance.
This is quite surprising, but it can be explained with three observations.
First, there are not users without ratings, so even a small amount of information is always available.
Second, long profiles include ratings that the user provided in different contexts, or in a long period of time, and since there is no information about the context where the evaluation is performed, it is difficult for the recommender to understand which ratings in the user profile are more relevant in the evaluation scenario.
Third, long profiles probably include popular items, forcing the recommenders to select niche alternatives that are harder to recommend\footnote{The sets of items to rank in validation and test sets do not overlap with the set of items rated by the respective user in any market or dataset.}.

The results obtained on different market combinations mainly confirm the results observed in Table \ref{tab:gbdt-validation-results}. Source markets s2 and s3 are beneficial for the accuracy on the t1 target market, while s1 enhances the performance of the model on target market t2.
Overall, it is interesting to notice that, while the datasets composed of only one target market obtain competitive performance, the best accuracy is achieved on datasets that include additional source markets, highlighting the importance of cross-market recommendation.

Finally, in Table \ref{tab:final-validation-results} we show the results, in terms of NDCG@10, obtained by the last stage ensembles of our model.
Both the ensembles reach higher ranking accuracy with respect to the ensembles of the second stage, demonstrating that merging the information coming from different combinations of datasets is beneficial for the model performance.
The LightGBM ensemble achieves the highest accuracy overall, and it represents the basis of the final submission of our team.
The scores ensemble, instead, was used to break ties in the GBDT solution, as we noticed that GBDT models tended to predict, given a user, the exact same scores for many items, a behavior that can be detrimental for the evaluation of the ranking accuracy.
This solution allowed our team to reach the 4th position in the final leaderboard.

\begin{table}[t]
\small
\setlength{\tabcolsep}{3pt}
\caption{NDCG@10 of the best hyperparameter configuration obtained on the validation set of each target market dataset of the Scores Combination and LightGBM last-level ensembles.}
\label{tab:final-validation-results}
\begin{tabular}{l|cccc|c|c} \toprule
\multirow{2}{*}{\thead{Target \\ Market}} & \multicolumn{5}{c|}{\thead{Scores Ensemble by Profile Length}} & \multicolumn{1}{c}{\multirow{2}{*}{\thead{LGBM\\Ens.}}} \\
 & \multicolumn{1}{c}{\thead{Short}} & \multicolumn{1}{c}{\thead{Quite\\Short}} & \multicolumn{1}{c}{\thead{Quite\\Long}} & \multicolumn{1}{c|}{\thead{Long}} & \multicolumn{1}{c|}{\thead{Avg}} &  \\ \midrule 
t1  &  0.97525 & 0.74789 & 0.65847 & 0.60235 & 0.71211 & \textbf{0.72325} \\
t2  &  0.94728 & 0.67463 & 0.56911 & 0.50388 & 0.61853 & \textbf{0.63471} \\
\bottomrule
\end{tabular}
\end{table}

%% file: tables/tab-gbdt-validation-results.tex
\begin{table}[t]
\footnotesize
\setlength{\tabcolsep}{1.8pt}
\caption{NDCG@10 of the best hyperparameter configuration of the GBDT dataset-level ensemble models obtained on the validation set for each dataset. Best dataset performance for each target market is in bold.}
\label{tab:gbdt-validation-results}
\begin{tabular}{l|l|ccc|ccc|c} \toprule
\multirow{2}{*}{\thead{Tgt\\Mkt}} & \multicolumn{1}{c|}{\multirow{2}{*}{\thead{Dataset}}} & \multicolumn{3}{c|}{\thead{LightGBM}} & \multicolumn{3}{c|}{\thead{XGBoost}} & \multicolumn{1}{c}{\thead{CatBoost*}} \\ 
 & & \multicolumn{1}{c}{\thead{Norm}} & \multicolumn{1}{c}{\thead{No\\Norm}} & \multicolumn{1}{c|}{\thead{Both}} & \multicolumn{1}{c}{\thead{Norm}} & \multicolumn{1}{c}{\thead{No\\Norm}} & \multicolumn{1}{c|}{\thead{Both}} & \multicolumn{1}{c}{\thead{Norm}} \\ \midrule
\multirow{16}{*}{t1} & s1-s2-s3-t1 & 0.70415 & 0.71391 & 0.71337 & 0.70201 & 0.70872 & 0.70734 & \textbf{0.71043} \\
 & s1-s2-s3-t1-t2 & 0.69878 & 0.70889 & 0.70565 & \textbf{0.70206} & 0.71081 & 0.70174 & 0.70284 \\
 & s1-s2-t1 & 0.70360 & 0.71269 & 0.71396 & 0.69866 & 0.71252 & 0.70993 & 0.69783 \\
 & s1-s2-t1-t2 & 0.70255 & 0.71061 & 0.71032 & 0.70068 & 0.70934 & 0.70881 & 0.69597 \\
 & s1-s3-t1 & 0.70319 & 0.71242 & 0.71190 & 0.69870 & \textbf{0.71425} & 0.70791 & 0.69691 \\
 & s1-s3-t1-t2 & 0.69964 & 0.71296 & 0.71058 & 0.70099 & 0.71337 & 0.70355 & 0.69827 \\
 & s1-t1 & 0.70040 & 0.70996 & 0.70731 & 0.69255 & 0.70849 & 0.70548 & 0.69533 \\
 & s1-t1-t2 & 0.69530 & 0.70854 & 0.70712 & 0.69505 & 0.70113 & 0.70169 & 0.69376 \\
 & s2-s3-t1 & \textbf{0.70526} & \textbf{0.71446} & \textbf{0.71405} & 0.70100 & 0.70770 & 0.70452 & 0.70112 \\
 & s2-s3-t1-t2 & 0.70168 & 0.71274 & 0.70801 & 0.70182 & 0.71401 & \textbf{0.71228} & 0.70214 \\
 & s2-t1 & 0.70036 & 0.70875 & 0.70540 & 0.69815 & 0.70336 & 0.70254 & 0.70479 \\
 & s2-t1-t2 & 0.70043 & 0.70718 & 0.70545 & 0.69838 & 0.70993 & 0.70307 & 0.69916 \\
 & s3-t1 & 0.70297 & 0.71163 & 0.70835 & 0.69754 & 0.70868 & 0.70521 & 0.69880 \\
 & s3-t1-t2 & 0.70340 & 0.70669 & 0.70494 & 0.70156 & 0.70968 & 0.70594 & 0.70076 \\
 & t1-t2 & 0.69914 & 0.70762 & 0.70468 & 0.69611 & 0.70342 & 0.70001 & 0.69854 \\
\cmidrule{2-9}
 & t1 & 0.70321 & 0.70605 & 0.70893 & 0.69681 & 0.70729 & 0.70454 & 0.69717 \\
\midrule
\multirow{16}{*}{t2} & s1-s2-s3-t1-t2 & 0.60983 & 0.62106 & 0.62178 & 0.60731 & 0.61894 & 0.62015 & 0.60629 \\
 & s1-s2-s3-t2 & 0.60841 & 0.62121 & 0.61704 & 0.60550 & 0.61549 & 0.61924 & 0.60640 \\
 & s1-s2-t1-t2 & \textbf{0.61071} & 0.61951 & 0.61769 & 0.60937 & 0.61724 & 0.61502 & 0.61013 \\
 & s1-s2-t2 & 0.60817 & 0.62315 & 0.61894 & \textbf{0.60947} & 0.61940 & 0.61593 & 0.60890 \\
 & s1-s3-t1-t2 & 0.60898 & 0.62025 & 0.62058 & 0.60713 & 0.61927 & 0.61996 & 0.60764 \\
 & s1-s3-t2 & 0.60878 & 0.62228 & 0.62128 & 0.60723 & \textbf{0.62203} & 0.61754 & 0.60689 \\
 & s1-t1-t2 & 0.61063 & 0.62254 & 0.62047 & 0.60820 & 0.62110 & \textbf{0.62133} & \textbf{0.61633} \\
 & s1-t2 & 0.60864 & \textbf{0.62703} & \textbf{0.62673} & 0.60474 & 0.62068 & 0.62006 & 0.60829 \\
 & s2-s3-t1-t2 & 0.60849 & 0.61632 & 0.61162 & 0.60521 & 0.61486 & 0.61252 & 0.601099 \\
 & s2-s3-t2 & 0.60586 & 0.61152 & 0.61146 & 0.60277 & 0.61276 & 0.61061 & 0.60004 \\
 & s2-t1-t2 & 0.60591 & 0.61263 & 0.61074 & 0.60476 & 0.61258 & 0.61082 & 0.60579 \\
 & s2-t2 & 0.60584 & 0.61223 & 0.61171 & 0.60312 & 0.60552 & 0.61038 & 0.60846 \\
 & s3-t1-t2 & 0.60689 & 0.61466 & 0.61257 & 0.60627 & 0.61415 & 0.60935 & 0.60590 \\
 & s3-t2 & 0.60871 & 0.61715 & 0.61352 & 0.60257 & 0.61624 & 0.61291 & 0.60630\\
 & t1-t2 & 0.60749 & 0.61338 & 0.61343 & 0.60310 & 0.60448 & 0.61209 & 0.60400 \\
\cmidrule{2-9}
 & t2 & 0.60979 & 0.61589 & 0.61555 & 0.60202 & 0.61298 & 0.61252 & 0.60505 \\
\bottomrule
\end{tabular}
\footnotesize{*Due to the limited availability of GPU computational resources, it was not possible to perform the hyperparameter optimization of CatBoost for all the normalization types.}
\end{table}

%% file: tables/tab-scores-validation-results.tex
\begin{table}[t]
\footnotesize
\setlength{\tabcolsep}{2pt}
\caption{NDCG@10 of the best hyperparameter configuration obtained on the validation set for each dataset of the Scores Combination dataset-level ensemble. Best performance among datasets for each target market is in bold.}
\label{tab:scores-validation-results}
\begin{tabular}{l|l|cccc|c} \toprule
\thead{Tgt\\Mkt} & \multicolumn{1}{c|}{\thead{Dataset}} & \multicolumn{1}{c}{\thead{Short}} & \multicolumn{1}{c}{\thead{Quite\\Short}} & \multicolumn{1}{c}{\thead{Quite\\Long}} & \multicolumn{1}{c|}{\thead{Long}} & \multicolumn{1}{c}{\thead{Avg}} \\ \midrule 
\multirow{16}{*}{t1}
& s1-s2-s3-t1 & 0.95652 & 0.73981 & 0.65213 & \textbf{0.59087} & 0.70397 \\
& s1-s2-s3-t1-t2 & 0.93818 & 0.74122 & 0.64961 & 0.57234 & 0.70202 \\
& s1-s2-t1 & 0.94726 & 0.73069 & 0.65195 & 0.58777 & 0.69766 \\
& s1-s2-t1-t2 & 0.95729 & 0.73318 & 0.64120 & 0.56736 & 0.69447 \\
& s1-s3-t1 & 0.94048 & \textbf{0.74505} & 0.63850 & 0.57107 & 0.70170 \\
& s1-s3-t1-t2 & \textbf{0.97525} & 0.73309 & 0.64124 & 0.57521 & 0.69546 \\
& s1-t1 & 0.95160 & 0.73760 & 0.63820 & 0.57014 & 0.69685 \\
& s1-t1-t2 & 0.95920 & 0.73820 & 0.63621 & 0.56212 & 0.69592 \\
& s2-s3-t1 & 0.95356 & 0.73770 & 0.65253 & 0.58818 & 0.70238 \\
& s2-s3-t1-t2 & 0.97024 & 0.74091 & \textbf{0.65264} & 0.58574 & \textbf{0.70433} \\
& s2-t1 & 0.95920 & 0.73048 & 0.64504 & 0.58679 & 0.69587 \\
& s2-t1-t2 & 0.95920 & 0.73545 & 0.64750 & 0.56453 & 0.69712 \\
& s3-t1 & \textbf{0.97525} & 0.73968 & 0.65258 & 0.57898 & 0.70281 \\
& s3-t1-t2 & 0.93478 & 0.73811 & 0.64349 & 0.56256 & 0.69744 \\
& t1-t2 & 0.95652 & 0.73929 & 0.64520 & 0.55725 & 0.69819 \\
\cmidrule{2-7}
& t1 & 0.97024 & 0.73725 & 0.65126 & 0.58237 & 0.70128 \\
\midrule
\multirow{16}{*}{t2}
& s1-s2-s3-t1-t2 & 0.92362 & 0.66463 & 0.55214 & 0.48346 & 0.60488 \\
& s1-s2-s3-t2 & 0.88478 & 0.65645 & 0.55685 & 0.49650 & 0.60361 \\
& s1-s2-t1-t2 & 0.91667 & 0.65901 & 0.55105 & 0.48285 & 0.60242 \\
& s1-s2-t2 & 0.92765 & 0.66233 & 0.55502 & 0.49703 & 0.60655 \\
& s1-s3-t1-t2 & 0.89726 & 0.66293 & 0.55766 & 0.49397 & 0.60708 \\
& s1-s3-t2 & 0.92765 & \textbf{0.66886} & 0.54512 & 0.49627 & 0.60721 \\
& s1-t1-t2 & 0.91867 & 0.66764 & 0.55939 & 0.48698 & 0.60916 \\
& s1-t2 & 0.92362 & 0.66769 & \textbf{0.56014} & \textbf{0.50027} & \textbf{0.61148} \\
& s2-s3-t1-t2 & 0.85331 & 0.66306 & 0.54768 & 0.48522 & 0.60282 \\
& s2-s3-t2 & 0.85714 & 0.65997 & 0.54848 & 0.47951 & 0.60046 \\
& s2-t1-t2 & 0.87585 & 0.66158 & 0.54337 & 0.48716 & 0.60112 \\
& s2-t2 & 0.89966 & 0.66050 & 0.54597 & 0.48322 & 0.60287 \\
& s3-t1-t2 & 0.89194 & 0.65540 & 0.55120 & 0.48080 & 0.59899 \\
& s3-t2 & \textbf{0.92857} & 0.66382 & 0.55181 & 0.48504 & 0.60459 \\
& t1-t2 & 0.88903 & 0.65632 & 0.54722 & 0.48861 & 0.59956 \\
\cmidrule{2-7}
& t2 & 0.90221 & 0.66200 & 0.55131 & 0.49294 & 0.60460 \\
\bottomrule
\end{tabular}
\end{table}

%% file: 10-conclusions.tex
\section{Conclusions}
\label{sec:conclusion}

In this work, we described the approach that lead our team PolimiRank to reach the 4th position in the WSDM Cup 2022 on cross-market recommendation.
Our solution is a multi-stage model where the scores predicted by state-of-the-art recommenders in the first stage are ensembled together, including other statistics mined from markets data, in two subsequent phases.
Ensembles are performed through a linear combination of recommenders' scores, and more powerful GBDT models optimized for the learning to rank task.
The results prove that exploiting information from different markets allows enhancing the ranking accuracy in a cross-market recommendation scenario.

%% file: main.bbl

\begin{thebibliography}{15}


\ifx \showCODEN    \undefined \def \showCODEN     #1{\unskip}     \fi
\ifx \showDOI      \undefined \def \showDOI       #1{#1}\fi
\ifx \showISBNx    \undefined \def \showISBNx     #1{\unskip}     \fi
\ifx \showISBNxiii \undefined \def \showISBNxiii  #1{\unskip}     \fi
\ifx \showISSN     \undefined \def \showISSN      #1{\unskip}     \fi
\ifx \showLCCN     \undefined \def \showLCCN      #1{\unskip}     \fi
\ifx \shownote     \undefined \def \shownote      #1{#1}          \fi
\ifx \showarticletitle \undefined \def \showarticletitle #1{#1}   \fi
\ifx \showURL      \undefined \def \showURL       {\relax}        \fi
\providecommand\bibfield[2]{#2}
\providecommand\bibinfo[2]{#2}
\providecommand\natexlab[1]{#1}
\providecommand\showeprint[2][]{arXiv:#2}

\bibitem[Bonab et~al\mbox{.}(2021)]%
        {cross-market-1}
\bibfield{author}{\bibinfo{person}{Hamed~R. Bonab}, \bibinfo{person}{Mohammad
  Aliannejadi}, \bibinfo{person}{Ali Vardasbi}, \bibinfo{person}{Evangelos
  Kanoulas}, {and} \bibinfo{person}{James Allan}.}
  \bibinfo{year}{2021}\natexlab{}.
\newblock \showarticletitle{Cross-Market Product Recommendation}. In
  \bibinfo{booktitle}{\emph{{CIKM} '21: The 30th {ACM} International Conference
  on Information and Knowledge Management, Virtual Event, Queensland,
  Australia, November 1 - 5, 2021}},
  \bibfield{editor}{\bibinfo{person}{Gianluca Demartini},
  \bibinfo{person}{Guido Zuccon}, \bibinfo{person}{J.~Shane Culpepper},
  \bibinfo{person}{Zi~Huang}, {and} \bibinfo{person}{Hanghang Tong}} (Eds.).
  \bibinfo{publisher}{{ACM}}, \bibinfo{pages}{110--119}.
\newblock
\urldef\tempurl%
\url{https://doi.org/10.1145/3459637.3482493}
\showDOI{\tempurl}


\bibitem[Carminati et~al\mbox{.}(2021)]%
        {recsys-challenge-1}
\bibfield{author}{\bibinfo{person}{Luca Carminati}, \bibinfo{person}{Giacomo
  Lodigiani}, \bibinfo{person}{Pietro Maldini}, \bibinfo{person}{Samuele Meta},
  \bibinfo{person}{Stiven Metaj}, \bibinfo{person}{Arcangelo Pisa},
  \bibinfo{person}{Alessandro Sanvito}, \bibinfo{person}{Mattia Surricchio},
  \bibinfo{person}{Fernando Benjam{\'{\i}}n~P{\'{e}}rez Maurera},
  \bibinfo{person}{Cesare Bernardis}, {and} \bibinfo{person}{Maurizio~Ferrari
  Dacrema}.} \bibinfo{year}{2021}\natexlab{}.
\newblock \showarticletitle{Lightweight and Scalable Model for Tweet
  Engagements Predictions in a Resource-constrained Environment}. In
  \bibinfo{booktitle}{\emph{RecSys Challenge 2021: Proceedings of the
  Recommender Systems Challenge 2021, Amsterdam, The Netherlands, 1 October
  2021}}. \bibinfo{publisher}{{ACM}}, \bibinfo{pages}{28--33}.
\newblock
\urldef\tempurl%
\url{https://doi.org/10.1145/3487572.3487597}
\showDOI{\tempurl}


\bibitem[Cooper et~al\mbox{.}(2014)]%
        {p3alpha}
\bibfield{author}{\bibinfo{person}{Colin Cooper}, \bibinfo{person}{Sang{-}Hyuk
  Lee}, \bibinfo{person}{Tomasz Radzik}, {and} \bibinfo{person}{Yiannis
  Siantos}.} \bibinfo{year}{2014}\natexlab{}.
\newblock \showarticletitle{Random walks in recommender systems: exact
  computation and simulations}. In \bibinfo{booktitle}{\emph{23rd International
  World Wide Web Conference, {WWW} '14, Seoul, Republic of Korea, April 7-11,
  2014, Companion Volume}}, \bibfield{editor}{\bibinfo{person}{Chin{-}Wan
  Chung}, \bibinfo{person}{Andrei~Z. Broder}, \bibinfo{person}{Kyuseok Shim},
  {and} \bibinfo{person}{Torsten Suel}} (Eds.). \bibinfo{publisher}{{ACM}},
  \bibinfo{pages}{811--816}.
\newblock
\urldef\tempurl%
\url{https://doi.org/10.1145/2567948.2579244}
\showDOI{\tempurl}


\bibitem[Cremonesi et~al\mbox{.}(2010)]%
        {puresvd}
\bibfield{author}{\bibinfo{person}{Paolo Cremonesi}, \bibinfo{person}{Yehuda
  Koren}, {and} \bibinfo{person}{Roberto Turrin}.}
  \bibinfo{year}{2010}\natexlab{}.
\newblock \showarticletitle{Performance of Recommender Algorithms on Top-n
  Recommendation Tasks}. In \bibinfo{booktitle}{\emph{Proceedings of the Fourth
  ACM Conference on Recommender Systems}} (Barcelona, Spain)
  \emph{(\bibinfo{series}{RecSys ’10})}. \bibinfo{publisher}{Association for
  Computing Machinery}, \bibinfo{address}{New York, NY, USA},
  \bibinfo{pages}{39–46}.
\newblock
\showISBNx{9781605589060}
\urldef\tempurl%
\url{https://doi.org/10.1145/1864708.1864721}
\showDOI{\tempurl}


\bibitem[Deotte et~al\mbox{.}(2021)]%
        {recsys-challenge-2}
\bibfield{author}{\bibinfo{person}{Chris Deotte}, \bibinfo{person}{Bo Liu},
  \bibinfo{person}{Benedikt Schifferer}, {and} \bibinfo{person}{Gilberto
  Titericz}.} \bibinfo{year}{2021}\natexlab{}.
\newblock \showarticletitle{{GPU} Accelerated Boosted Trees and Deep Neural
  Networks for Better Recommender Systems}. In \bibinfo{booktitle}{\emph{RecSys
  Challenge 2021: Proceedings of the Recommender Systems Challenge 2021,
  Amsterdam, The Netherlands, 1 October 2021}}. \bibinfo{publisher}{{ACM}},
  \bibinfo{pages}{7--14}.
\newblock
\urldef\tempurl%
\url{https://doi.org/10.1145/3487572.3487605}
\showDOI{\tempurl}


\bibitem[Hu et~al\mbox{.}(2008)]%
        {ials}
\bibfield{author}{\bibinfo{person}{Yifan Hu}, \bibinfo{person}{Yehuda Koren},
  {and} \bibinfo{person}{Chris Volinsky}.} \bibinfo{year}{2008}\natexlab{}.
\newblock \showarticletitle{Collaborative Filtering for Implicit Feedback
  Datasets}. In \bibinfo{booktitle}{\emph{Proceedings of the 8th {IEEE}
  International Conference on Data Mining {(ICDM} 2008), December 15-19, 2008,
  Pisa, Italy}}. \bibinfo{publisher}{{IEEE} Computer Society},
  \bibinfo{pages}{263--272}.
\newblock
\urldef\tempurl%
\url{https://doi.org/10.1109/ICDM.2008.22}
\showDOI{\tempurl}


\bibitem[Liang et~al\mbox{.}(2018)]%
        {multvae}
\bibfield{author}{\bibinfo{person}{Dawen Liang}, \bibinfo{person}{Rahul~G.
  Krishnan}, \bibinfo{person}{Matthew~D. Hoffman}, {and} \bibinfo{person}{Tony
  Jebara}.} \bibinfo{year}{2018}\natexlab{}.
\newblock \showarticletitle{Variational Autoencoders for Collaborative
  Filtering}. In \bibinfo{booktitle}{\emph{Proceedings of the 2018 World Wide
  Web Conference on World Wide Web, {WWW} 2018, Lyon, France, April 23-27,
  2018}}, \bibfield{editor}{\bibinfo{person}{Pierre{-}Antoine Champin},
  \bibinfo{person}{Fabien Gandon}, \bibinfo{person}{Mounia Lalmas}, {and}
  \bibinfo{person}{Panagiotis~G. Ipeirotis}} (Eds.).
  \bibinfo{publisher}{{ACM}}, \bibinfo{pages}{689--698}.
\newblock
\urldef\tempurl%
\url{https://doi.org/10.1145/3178876.3186150}
\showDOI{\tempurl}


\bibitem[Ning et~al\mbox{.}(2015)]%
        {neighborhood-based}
\bibfield{author}{\bibinfo{person}{Xia Ning}, \bibinfo{person}{Christian
  Desrosiers}, {and} \bibinfo{person}{George Karypis}.}
  \bibinfo{year}{2015}\natexlab{}.
\newblock \showarticletitle{A Comprehensive Survey of Neighborhood-Based
  Recommendation Methods}.
\newblock In \bibinfo{booktitle}{\emph{Recommender Systems Handbook}},
  \bibfield{editor}{\bibinfo{person}{Francesco Ricci}, \bibinfo{person}{Lior
  Rokach}, {and} \bibinfo{person}{Bracha Shapira}} (Eds.).
  \bibinfo{publisher}{Springer}, \bibinfo{pages}{37--76}.
\newblock
\urldef\tempurl%
\url{https://doi.org/10.1007/978-1-4899-7637-6\_2}
\showDOI{\tempurl}


\bibitem[Ning and Karypis(2011)]%
        {slim}
\bibfield{author}{\bibinfo{person}{Xia Ning} {and} \bibinfo{person}{George
  Karypis}.} \bibinfo{year}{2011}\natexlab{}.
\newblock \showarticletitle{{SLIM:} Sparse Linear Methods for Top-N Recommender
  Systems}. In \bibinfo{booktitle}{\emph{11th {IEEE} International Conference
  on Data Mining, {ICDM} 2011, Vancouver, BC, Canada, December 11-14, 2011}},
  \bibfield{editor}{\bibinfo{person}{Diane~J. Cook}, \bibinfo{person}{Jian
  Pei}, \bibinfo{person}{Wei Wang}, \bibinfo{person}{Osmar~R. Za{\"{\i}}ane},
  {and} \bibinfo{person}{Xindong Wu}} (Eds.). \bibinfo{publisher}{{IEEE}
  Computer Society}, \bibinfo{pages}{497--506}.
\newblock
\urldef\tempurl%
\url{https://doi.org/10.1109/ICDM.2011.134}
\showDOI{\tempurl}


\bibitem[Paudel et~al\mbox{.}(2017)]%
        {rp3beta}
\bibfield{author}{\bibinfo{person}{Bibek Paudel}, \bibinfo{person}{Fabian
  Christoffel}, \bibinfo{person}{Chris Newell}, {and} \bibinfo{person}{Abraham
  Bernstein}.} \bibinfo{year}{2017}\natexlab{}.
\newblock \showarticletitle{Updatable, Accurate, Diverse, and Scalable
  Recommendations for Interactive Applications}.
\newblock \bibinfo{journal}{\emph{{ACM} Trans. Interact. Intell. Syst.}}
  \bibinfo{volume}{7}, \bibinfo{number}{1} (\bibinfo{year}{2017}),
  \bibinfo{pages}{1:1--1:34}.
\newblock
\urldef\tempurl%
\url{https://doi.org/10.1145/2955101}
\showDOI{\tempurl}


\bibitem[Roitero et~al\mbox{.}(2020)]%
        {cross-market-2}
\bibfield{author}{\bibinfo{person}{Kevin Roitero}, \bibinfo{person}{Ben
  Carterette}, \bibinfo{person}{Rishabh Mehrotra}, {and}
  \bibinfo{person}{Mounia Lalmas}.} \bibinfo{year}{2020}\natexlab{}.
\newblock \showarticletitle{Leveraging Behavioral Heterogeneity Across Markets
  for Cross-Market Training of Recommender Systems}. In
  \bibinfo{booktitle}{\emph{Companion of The 2020 Web Conference 2020, Taipei,
  Taiwan, April 20-24, 2020}}, \bibfield{editor}{\bibinfo{person}{Amal
  El~Fallah Seghrouchni}, \bibinfo{person}{Gita Sukthankar},
  \bibinfo{person}{Tie{-}Yan Liu}, {and} \bibinfo{person}{Maarten van Steen}}
  (Eds.). \bibinfo{publisher}{{ACM} / {IW3C2}}, \bibinfo{pages}{694--702}.
\newblock
\urldef\tempurl%
\url{https://doi.org/10.1145/3366424.3384362}
\showDOI{\tempurl}


\bibitem[Shenbin et~al\mbox{.}(2020)]%
        {recvae}
\bibfield{author}{\bibinfo{person}{Ilya Shenbin}, \bibinfo{person}{Anton
  Alekseev}, \bibinfo{person}{Elena Tutubalina}, \bibinfo{person}{Valentin
  Malykh}, {and} \bibinfo{person}{Sergey~I. Nikolenko}.}
  \bibinfo{year}{2020}\natexlab{}.
\newblock \showarticletitle{RecVAE: {A} New Variational Autoencoder for Top-N
  Recommendations with Implicit Feedback}. In \bibinfo{booktitle}{\emph{{WSDM}
  '20: The Thirteenth {ACM} International Conference on Web Search and Data
  Mining, Houston, TX, USA, February 3-7, 2020}},
  \bibfield{editor}{\bibinfo{person}{James Caverlee},
  \bibinfo{person}{Xia~(Ben) Hu}, \bibinfo{person}{Mounia Lalmas}, {and}
  \bibinfo{person}{Wei Wang}} (Eds.). \bibinfo{publisher}{{ACM}},
  \bibinfo{pages}{528--536}.
\newblock
\urldef\tempurl%
\url{https://doi.org/10.1145/3336191.3371831}
\showDOI{\tempurl}


\bibitem[Steck(2019)]%
        {easer}
\bibfield{author}{\bibinfo{person}{Harald Steck}.}
  \bibinfo{year}{2019}\natexlab{}.
\newblock \showarticletitle{Embarrassingly Shallow Autoencoders for Sparse
  Data}. In \bibinfo{booktitle}{\emph{The World Wide Web Conference, {WWW}
  2019, San Francisco, CA, USA, May 13-17, 2019}},
  \bibfield{editor}{\bibinfo{person}{Ling Liu}, \bibinfo{person}{Ryen~W.
  White}, \bibinfo{person}{Amin Mantrach}, \bibinfo{person}{Fabrizio
  Silvestri}, \bibinfo{person}{Julian~J. McAuley}, \bibinfo{person}{Ricardo
  Baeza{-}Yates}, {and} \bibinfo{person}{Leila Zia}} (Eds.).
  \bibinfo{publisher}{{ACM}}, \bibinfo{pages}{3251--3257}.
\newblock
\urldef\tempurl%
\url{https://doi.org/10.1145/3308558.3313710}
\showDOI{\tempurl}


\bibitem[Weston et~al\mbox{.}(2011)]%
        {warp}
\bibfield{author}{\bibinfo{person}{Jason Weston}, \bibinfo{person}{Samy
  Bengio}, {and} \bibinfo{person}{Nicolas Usunier}.}
  \bibinfo{year}{2011}\natexlab{}.
\newblock \showarticletitle{{WSABIE:} Scaling Up to Large Vocabulary Image
  Annotation}. In \bibinfo{booktitle}{\emph{{IJCAI} 2011, Proceedings of the
  22nd International Joint Conference on Artificial Intelligence, Barcelona,
  Catalonia, Spain, July 16-22, 2011}}, \bibfield{editor}{\bibinfo{person}{Toby
  Walsh}} (Ed.). \bibinfo{publisher}{{IJCAI/AAAI}},
  \bibinfo{pages}{2764--2770}.
\newblock
\urldef\tempurl%
\url{https://doi.org/10.5591/978-1-57735-516-8/IJCAI11-460}
\showDOI{\tempurl}


\bibitem[Weston et~al\mbox{.}(2013)]%
        {kos-warp}
\bibfield{author}{\bibinfo{person}{Jason Weston}, \bibinfo{person}{Hector Yee},
  {and} \bibinfo{person}{Ron~J. Weiss}.} \bibinfo{year}{2013}\natexlab{}.
\newblock \showarticletitle{Learning to rank recommendations with the k-order
  statistic loss}. In \bibinfo{booktitle}{\emph{Seventh {ACM} Conference on
  Recommender Systems, RecSys '13, Hong Kong, China, October 12-16, 2013}},
  \bibfield{editor}{\bibinfo{person}{Qiang Yang}, \bibinfo{person}{Irwin King},
  \bibinfo{person}{Qing Li}, \bibinfo{person}{Pearl Pu}, {and}
  \bibinfo{person}{George Karypis}} (Eds.). \bibinfo{publisher}{{ACM}},
  \bibinfo{pages}{245--248}.
\newblock
\urldef\tempurl%
\url{https://doi.org/10.1145/2507157.2507210}
\showDOI{\tempurl}


\end{thebibliography}
